\documentclass[prl,aps,a4paper,twocolumn, floatfix]{revtex4-1}
\usepackage[utf8x]{inputenc}
\usepackage{graphicx}
\pdfoutput=1
\usepackage{latexsym}
\usepackage{amsmath}
\usepackage{amsfonts}
\usepackage{amssymb}
\usepackage{amsthm}
\usepackage[english]{babel}
\usepackage[caption=false]{subfig}
\usepackage[nouppercase]{scrpage2}
\usepackage{color}
\usepackage[colorlinks=true,bookmarks=false,pdfstartview={XYZ null null 1}]{hyperref} 
\usepackage[all]{hypcap} 
\usepackage{cleveref}
\newcommand{\moly}{MoS$_{2}\,$}
\begin{document}
\title{\Large
  {\bf Field-induced dissociation of excitons in two-dimensional MoS$_{2}$/hBN heterostructures}
  \vspace{0.5cm}}
\author{Sten Haastrup}
\affiliation{Center for Nanostructured Graphene (CNG), Department of Physics, Technical University of Denmark}
\affiliation{Center for Atomic-Scale Materials Design (CAMD), Department of Physics, Technical University of Denmark}

\author{Simone Latini}
\affiliation{Center for Nanostructured Graphene (CNG), Department of Physics, Technical University of Denmark}
\affiliation{Center for Atomic-Scale Materials Design (CAMD), Department of Physics, Technical University of Denmark}

\author{Kirill Bolotin}
\affiliation{Department of Physics and Astronomy, Vanderbilt University, Nashville, Tennessee 37235, United States}
\author{Kristian S. Thygesen}
\affiliation{Center for Nanostructured Graphene (CNG), Department of Physics, Technical University of Denmark}
\affiliation{Center for Atomic-Scale Materials Design (CAMD), Department of Physics, Technical University of Denmark}

\begin{abstract}
Atomically thin semi-conductors are characterized by strongly bound excitons which govern the optical properties of the materials below and near the band edge.
Efficient conversion of photons into electrical current requires, as a first step, the dissociation of the exciton into free electrons and holes.
Here we calculate the dissociation rates of excitons in monolayer MoS$_2$ as a function of an applied in-plane electric field.
The dissociation rates are obtained as the inverse lifetime of the resonant states of a two-dimensional Hydrogenic Hamiltonian which describes the exciton within the Mott-Wannier model.
The resonances are computed using complex scaling, and the effective masses and screened electron-hole interaction defining the Hydrogenic Hamiltonian are computed from first-principles.
For field strengths above 0.1 V/nm the dissociation lifetime is shorter than 1 picosecond, which is shorter than the lifetime of other, competing, decay mechanisms.
Interestingly, encapsulation of the \moly layer in just two layers of hBN, enhances the dissociation rate by around one order of magnitude due to the increased screening showing that dielectric engineering is an effective way to control exciton lifetimes in two-dimensional materials.   
\end{abstract}

\maketitle

\vspace{1cm}
Two-dimensional (2D) semiconductors, such as single- and few layer transition metal dichalcogenides, are presently being intensively researched due to their extraordinary electronic and optical properties which include strong light-matter interactions, spin-valley coupling, and easily tuneable electronic states\cite{Komsa2012, Wang2012, Fai2010, Ugeda2014, Keliang2014, Jariwala2014, Bernardi2013, Ross2014, Mak2012}.
One of the hallmarks of the 2D semiconductors is the presence of strongly bound excitons with binding energies reaching up to 30\% of the band gap.
These large binding energies are mainly a result of the reduced dielectric screening in two dimensions\cite{latini2015a,Hybertsen2013,Cudazzo2010,Cudazzo2011,Pulci2012}.
While such strongly bound excitons are highly interesting from a fundamental point of view (for example in the context of Bose-Einstein condensates\cite{Fogler2014}) they are problematic for many of the envisioned applications of 2D materials such as photodetectors and solar cells which rely on efficient conversion of photons into electrical currents.
This is because the strong attraction between the electron and the hole makes it difficult to dissociate the excitons into free carriers.

Photocurrent measurements on suspended MoS$_2$ samples have found that the photocurrent produced by below-band gap photons is strongly dependent on the applied voltage indicating that the electric field plays an important role in the generation of free carriers\cite{bolotin2014}.
One way of increasing the photoresponse, could be to embed the active 2D material into a van der Waals heterostructure\cite{Terrones2013,Britnell2013,Geim2013}, and thereby enhance the screening of the electron-hole interaction without altering the band structure of the material.
While it has been demonstrated that this strategy can indeed be used to control the exciton binding energy, the influence of environmental screening on the exciton dissociation has not been previously studied.

In general, rigorous calculations of exciton binding energies require a many-body approach such as the Bethe-Salpeter Equation (BSE) which directly solves for the (real) poles of the interacting response function which correspond to the neutral excitation energies of the system\cite{Strinati1984,Onida2002}.
Such calculations are  computationally demanding and typically only used to study excitations from the ground state, i.e. not in the presence of external fields.
We mention, however, that the BSE has been used to study field induced exciton dissociation in carbon nanotubes by fitting the BSE absorption spectrum to the Fano line shape\cite{perebeinos2007a}.
In this work we take a different approach using that, under certain simplifying circumstances, the calculation of the many-body excitonic state can be reformulated as an effective hydrogenic Hamiltonian whose eigenvalues and eigenstates represent the exciton binding energies and the envelope wave function describing the relative electron-hole motion, respectively.
This is the so-called Mott-Wannier model which has been instrumental in the description of excitons in inorganic bulk semiconductors.
A 2D version of the Mott-Wannier model has recently been shown to yield exciton binding energies in good agreement with BSE calculations and experiments for both freestanding\cite{latini2015a, olsen2016a, Hybertsen2013,Cudazzo2011,Pulci2012} and supported\cite{latini2015a, olsen2016a, Andersen2015} transition metal dichalcogenide layers.
The dissociation rate of the exciton is then obtained by complex scaling which is a formally exact technique to compute resonance energies and life times.
By employing a recently developed quantum-classical method for calculating the dielectric function of general van der Waals heterostructures, we predict the effect of embedding the MoS$_2$ in hBN on the screened electron-hole interaction and exciton dissociation rate.  

When an in-plane, constant electric field is applied to an exciton, it will eventually decay into a free electron and hole.
This effect belongs to a class first studied by Keldysh\cite{keldysh1957} and Franz\cite{franz1958}, who looked at how the optical properties of semiconductors change in the presence of a static electric field.
The application of a constant electric field changes the exciton from a bound state to a resonance with a finite lifetime equal to the inverse dissociation rate.
The literature on resonances in quantum physics is vast, and we shall not go into the topic here but simply mention a few few important facts.
First, it should be understood that even the definition of a resonance is non-trivial.
The reason for this is clear from Howland’s Razor which states that no satisfactory definition of a resonance can depend only on the structure of a single operator on an abstract Hilbert space\cite{simon1978a}.
To illustrate the content of the statement consider the Stark effect in hydrogen: Let $\hat{H}(\epsilon)=-\Delta - 1/r +\epsilon x$.
It can be shown that $\hat{H}(\epsilon)$ is unitarily equivalent to $\hat{H}(\epsilon’)$ for all nonzero $\epsilon,\epsilon’$.
Since we expect the properties of the resonances, and in particular their lifetime, to depend on field strength, $\epsilon$, this example shows that the resonance cannot be viewed as a property of the operator $\hat{H}(\epsilon)$ alone.
Instead the notion of resonance is only meaningful when the real space geometry of the given system and relevant boundary conditions on the wave functions are considered.
There are generally two approaches used to compute resonances.
The so-called indirect methods identify resonances as the poles of the scattering amplitude analytically extended to the complex energy plane\cite{taylor1972a}, while the direct methods obtain the resonance states directly as eigenstates of a complex scaled non-hermitian Hamiltonian\cite{balslev1971,aguilar1971a}. In this work we shall use the latter approach.

To describe excitons in a 2D semiconductor we use a Mott-Wannier model of the form
\begin{equation}
  \label{eq:MW}
  \left[-\frac{\nabla_{2D}^2}{2\mu_{ex}}+W({\bf r})\right] F({\bf
    r}_\parallel) = E_bF({\bf r}),
\end{equation}
where $\mu_{ex}$ is the exciton effective mass, $\mu_{ex}^{-1}=m_{e}^{-1}+m_{h}^{-1}$, $W$ is the screened electron-hole interaction, and $E_b$ denotes the exciton binding energy.
Based on density functional theory (DFT) band structure calculations using the local density approximation (LDA) we obtain an exciton effective mass for MoS$_2$ of 0.27$m_\mathrm{e}$.
The screened electron-hole interaction is obtained as the inverse Fourier transform of $[\epsilon_{2D}(q)q]^{-1}$, where $\epsilon_{2D}(q)$ is the static dielectric function of the 2D material and $1/q$ is the 2D Fourier transform of $1/r$.
For small $q$ we can approximate epsilon as a linear function of $q$\cite{Cudazzo2010, Cudazzo2011, Pulci2012, Hybertsen2013}, so that
\begin{equation}\label{eq:eps}
\epsilon_{2D}({\bf q}) = 1+2\pi\alpha q,
\end{equation}
with $\alpha$ being the polarizability of the material. An analytical expression can then be obtained for the screened electron-hole interaction\cite{Cudazzo2010}:
\begin{equation}
 W({\bf r})= \frac{1}{4\alpha} \left[Y_0(x)-H_0(x)\right]_{x=r/2\pi\alpha},
 \label{eq:W}
\end{equation}
where $Y_0$ is a Bessel function of the second kind, and $H_0$ is a Struve function.
For later use we note that both of these functions are meromorphic, and readily admit complex arguments.
The expression (\ref{eq:W}) for the screened interaction relies on a first order expansion of $\epsilon_{2D}(q)$ around $q=0$, and thus the results obtained from the Mott-Wannier model could be questioned.
However, the validity of this approximation has been demonstrated for a number of freestanding 2D semiconductors\cite{Hybertsen2013,Cudazzo2011,Pulci2012} and recently also for MoS$_2$ embedded in a few layers of hBN\cite{latini2015a}.
As a rule of thumb, the linear screening approximation Eq. (\ref{eq:eps}) remains valid for intra-layer excitons in van der Waals heterostructures as long as the in-plane exciton radius is large compared to the thickness of the heterostructure\cite{latini2015a}.
For thicker slabs, the linear approximation breaks down and the fully $q$-dependent $\epsilon_{2D}(q)$ must be used to obtain $W(r)$.
For details on how we calculate the dielectric functions of 2D layers and heterostructures we refer to Ref. \onlinecite{Andersen2015}.

\begin{figure}[b]
 \includegraphics[width=0.45\textwidth]{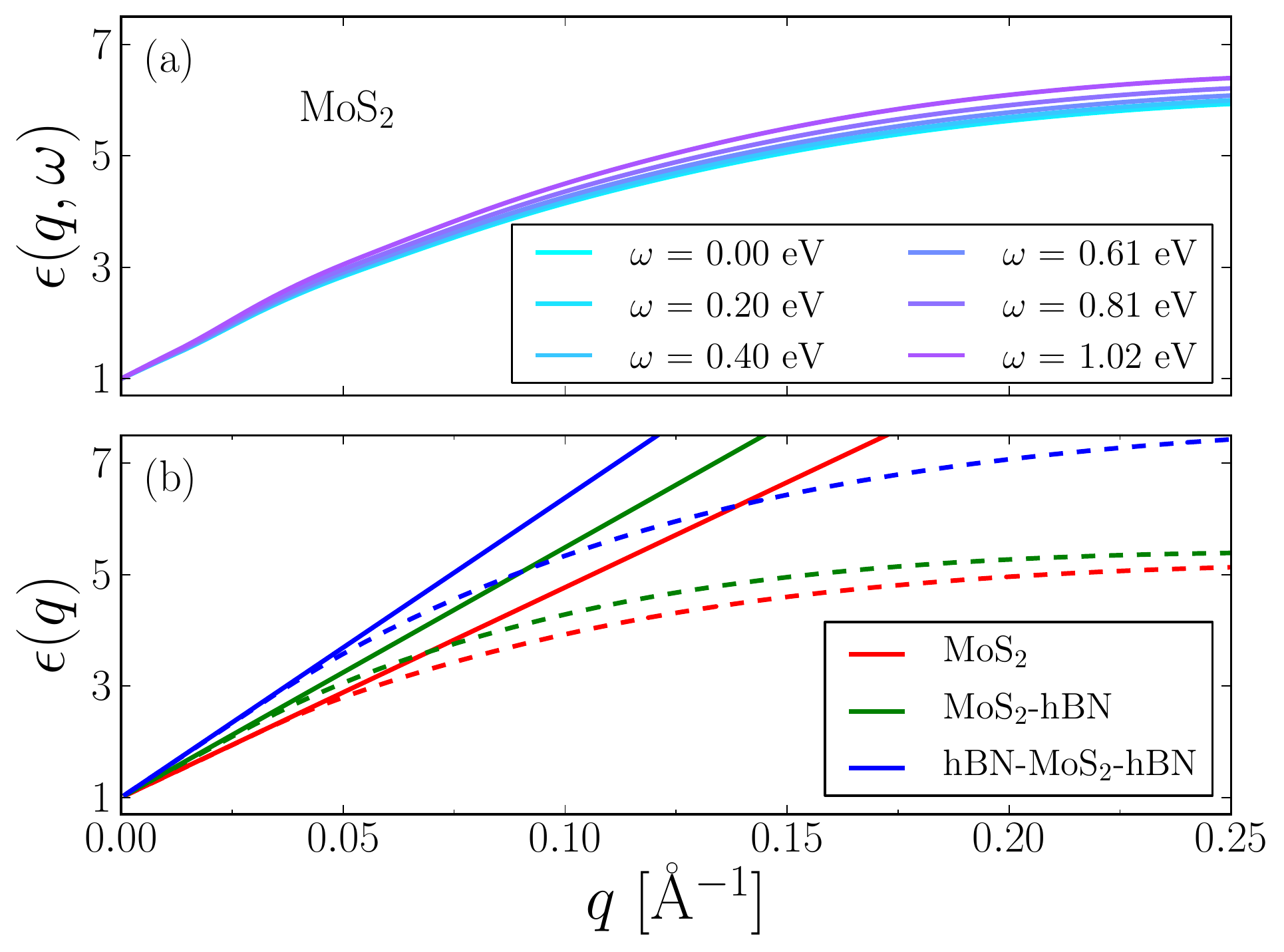}
 \caption{
(a) The effective dielectric function for isolated \moly as a function of $q$ for different values of $\omega$.
(b) The effective dielectric function (dashed line) for the three different MoS$_2$-hBN heterostructure configurations.
The linear approximation to the dielectric function is shown by solid lines.
}
 \label{fig:VdW_eps}
\end{figure}

\begin{figure*}[htb]
\subfloat{
      \includegraphics[height=0.28\textheight]{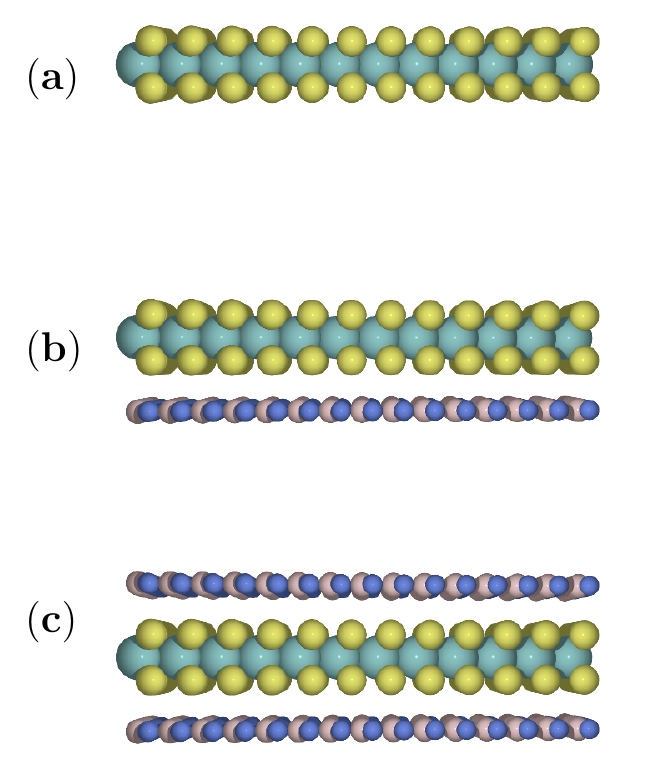}}
\hfill
\subfloat{
 \includegraphics[height=0.28\textheight,trim={0 1.15cm 0 0},clip]{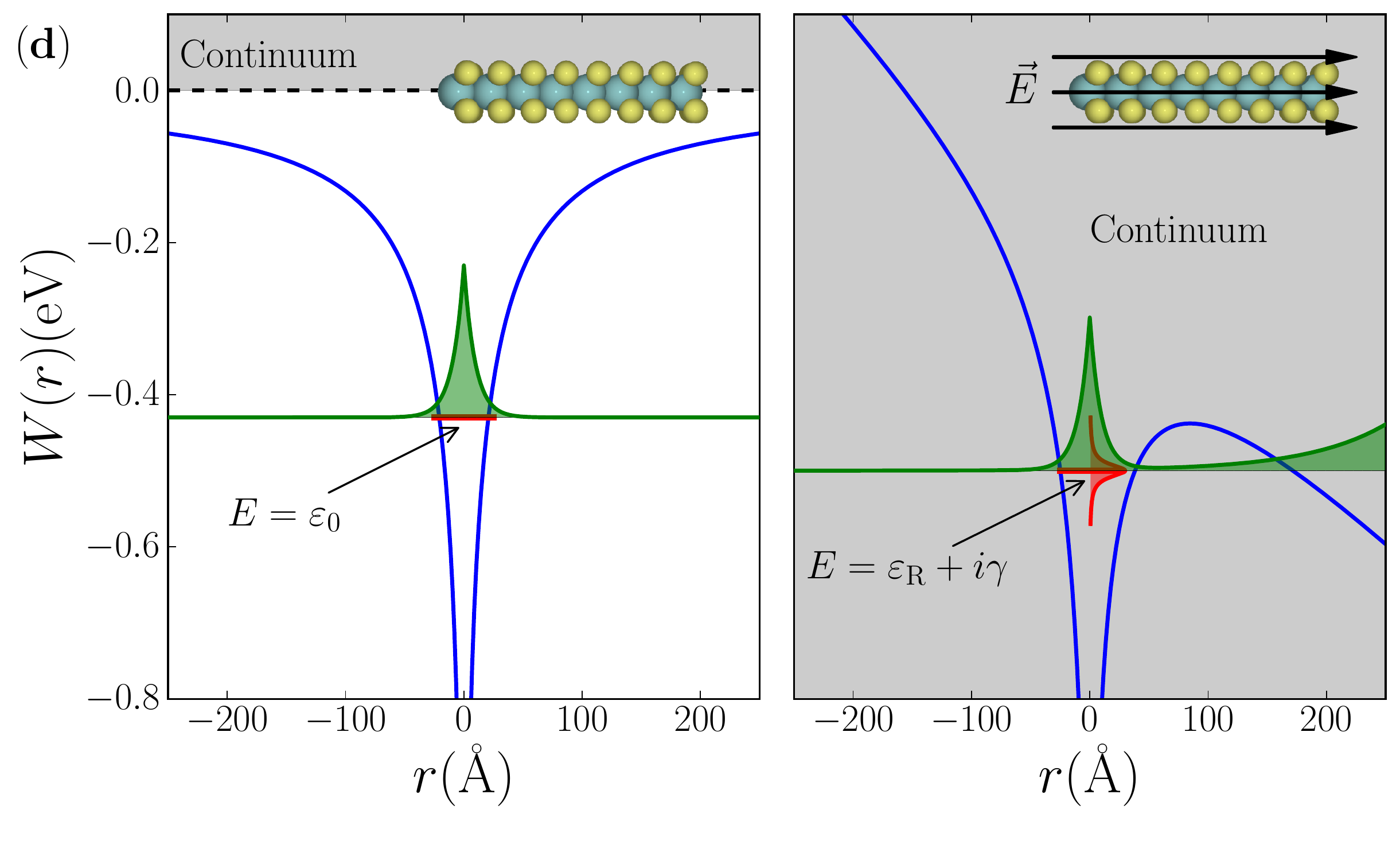}}
\caption{
\textbf{(a)-(c)}: The three different structures considered in this work: isolated \moly, \moly on a single layer of hBN, and \moly sandwiched between two hBN layers.
\textbf{(d)}:  Illustration of the Mott-Wannier model for monolayer MoS$_2$ in the absence (left) and presence (right) of an in-plane constant electric field.
The exciton potential is shown in blue, the exciton wave function is sketched in green, and the energy is shown in red.
When an electric field is applied, the energy of the exciton shifts down and the sharp energy peak is broadened due to the coupling to the continuum of states.\label{fig:physics}}
\end{figure*}

In \cref{fig:VdW_eps}, we show the dielectric functions, $\epsilon_{2D}(q)$, of an MoS$_2$ layer in the three different configurations shown in Fig.  \ref{fig:physics}(a)-(c).
The linear approximations to $\epsilon_{2D}(q)$ are shown by full lines.
We also show the dielectric function of isolated \moly for different values of $\omega$.
For values of $\omega$ which are below the exciton binding energy, the dielectric function is very similar to the static one, justifying our use of the latter.
\begin{table}[t]
 \label{tab:alpha}
 \caption{Calculated values for the polarisability ($\alpha$) and exciton binding energy ($E_b$) for single-layer MoS$_2$ in the three configurations shown in Fig. 2 (a)-(c).}
 \resizebox{0.3\textwidth}{!}{
   \begin{tabular}{ccccc}
     \hline
     Material & $\alpha$ (a.u.) & $E_b$ (eV) \\ \hline
     MoS$_2$ & 11.1 & 0.62 \\
     MoS$_2$@hBN & 13.0 & 0.55 \\
     hBN@MoS$_2$@hBN & 16.1  & 0.47  \\
     \hline
 \end{tabular}}
\end{table}
As expected, embedding of the MoS$_2$ layer in hBN leads to an increase in the screening which reduces the binding energy of the exciton, see \cref{tab:alpha}.

Once an in-plane constant electric field is applied to the system, the bound states of the Mott-Wannier Hamiltonian become metastable.
The situation is illustrated in Figure \ref{fig:physics}(d).
Within the so-called direct methods, a resonance is defined as an eigenstate of the Hamiltonian under the boundary condition that only outgoing waves exist outside the scattering region.
Such an eigenstate must necessarily have a complex eigenvalue, $E=\varepsilon_0-i\gamma$, and a wave function that adopts the asymptotic form $e^{\pm iKx}$ for $x\to \pm\infty$ (focusing on the one-dimensional case for simplicity) where $K=k-i\kappa$ with $k>0$ (an outgoing wave) and $\kappa>0$.
The latter condition implies that the wave function increases exponentially away from the scattering region.
The decay rate of the resonance state, evaluated as the rate of decay of the probability for finding the particle in any finite region of space, is given by $\gamma=k\kappa$.
It can be shown that the resonance eigenvalue, $E$, is a pole of the analytically continued scattering matrix\cite{hatano2008a}.

To compute the resonance, one could in principle solve the Schrödinger equation with the appropriate boundary conditions.
In practice, however, it is more convenient to perform a “complex scaling” of the Hamiltonian, whereby the coordinate $r \to e^{i\theta}r$ and $\nabla \to e^{-i\theta}\nabla$, and then solve for the eigenstates of the resulting (non-hermitian) operator, $\hat{H}_\theta$, with the more standard zero boundary conditions.
For $\theta>\tan^{-1}(\gamma/k)$, the complex scaled resonance wave function (that is the wavefunction analytically continued to the complex plane and then evaluated on the line $re^{i\theta}$) is an eigenstate of $\hat{H}_\theta$ of eigenvalue $E$, but now decaying exponentially as $r\to \pm \infty$.
The resonances thus appear as isolated complex eigenvalues of $\hat{H}_\theta$ with energy independent of $\theta$ and square integrable wave function\cite{reinhardt1982a}.
The complex scaled wave functions of the bound states remain exponentially decaying eigenstates of $\hat{H}_\theta$ with real eigenvalues\cite{balslev1971}.
The unbound continuum states have a different behavior: If the potentials involved are localized, the asymptotic form of these states as $r \to \infty$ is $e^{ikr}$, with $k,r \in \mathbb R$.
They are thus finite at infinity, but nonnormalizable.
If this is to remain true after the complex scaling is performed, the transformation $r \to r e^{i\theta}$ must be accompanied by the transformation $k \to k e^{-i\theta}$.
As the energy of a plane wave is proportional to $k^2$, the complex scaling operation results in the energy of the continuum states rotating into the complex plane at an angle of $2\theta$. This is also true for the Coulomb potential, despite its long-ranged nature\cite{reinhardt1982a}.

\begin{figure}
 \centering
 \includegraphics[width=\columnwidth]{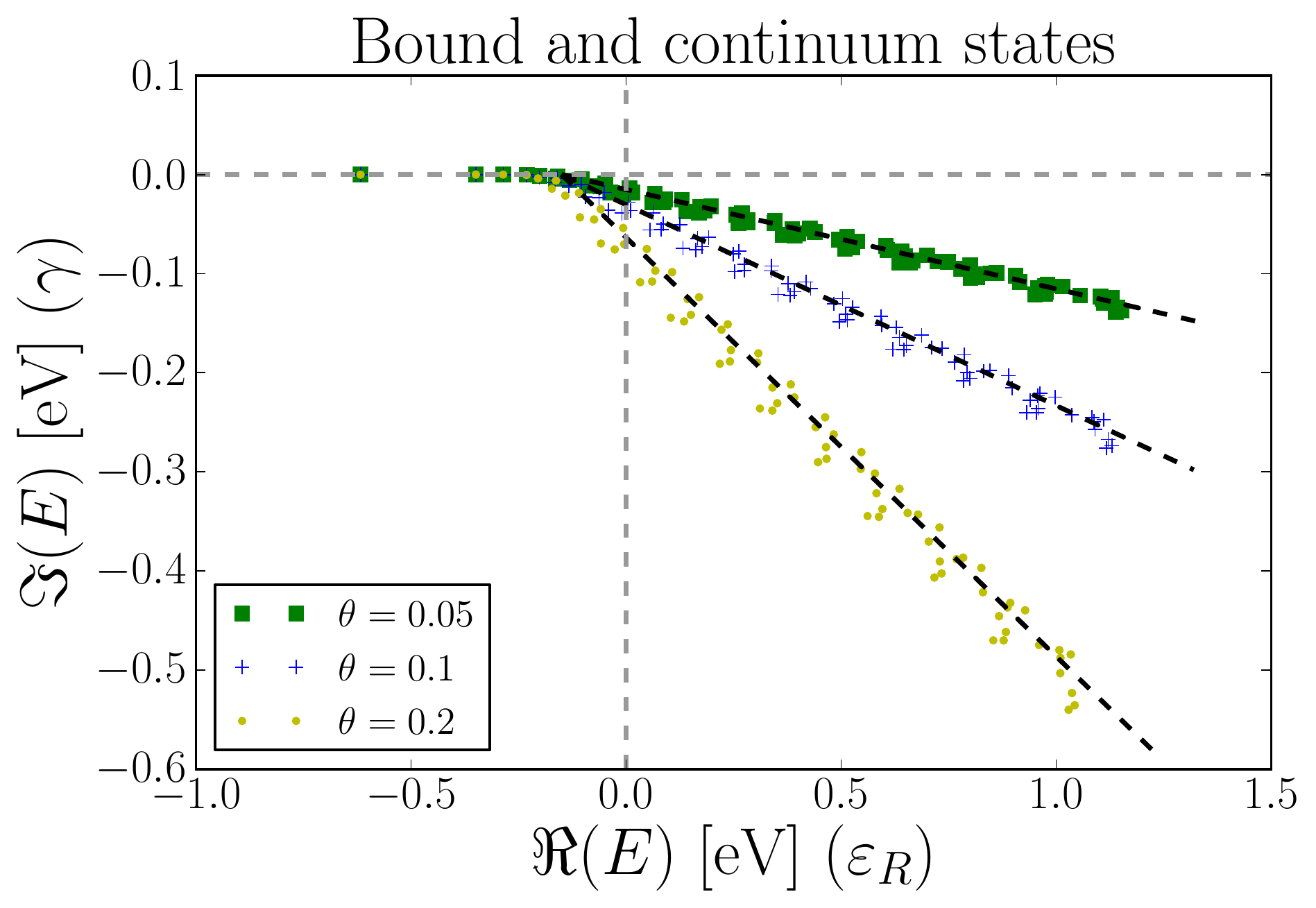}
 \caption{ \label{fig:argand}
   The different behaviour of bound and continuum states under complex scaling, for the potential corresponding to isolated \moly.
   The black dashed lines start at -0.15 eV and have been rotated into the complex plane by $-2\theta$ for each of the complex scaling angles.
   The continuum starts at -0.15 eV because of the finite size of the simulation box.
 }
\end{figure}

On figure \ref{fig:argand} we show an example of the spectrum of the complex-scaled exciton Hamiltonian for isolated \moly and zero field, for different values of $\theta$.
The two classes of states, bound and unbound, can clearly be distinguished; for zero field there are no resonances.

In order to apply the complex scaling procedure to a single-particle Hamiltonian, a number of requirements must be fulfilled by the potential, $V$.
First, in the original derivation of the complex scaling technique, the potential $\hat{V}(\mathbf r)$ should be \emph{dilatation analytic}\cite{aguilar1971a}.
This is the case for the Coulomb potential, and for many other potentials which have bound states, such as Yukawa potentials\cite{simon1973}.
A constant electric field is not dilatation analytic, but it has been proven that the technique works nonetheless\cite{herbst1978a,reinhardt2009a,herbst1979a}.
Secondly, most potentials of interest for physical systems are known only on the real axis.
In order to perform the complex scaling operation, it must be possible to find the analytic continuation of the potential in the complex plane.
This is the case for the potential considered here.
We mention that a full first-principles implementation of the complex scaling was recently reported and applied to the problem of Stark ionization of simple atoms\cite{larsen2013a}.

The 2D eigenvalue problem for the complex scaled Hamiltonian is solved on a real space grid using radial coordinates.
In order to converge the exciton energies, a large simulation cell is needed - significantly larger than the exciton radius, which is around 10 Å for all of the systems considered.
As the screened potential has a logarithmic singularity at the origin while being virtually flat at the edge of the simlation cell, a nonlinear grid is used which allows us to perform simulations in a disk of radius 250 Å.
The Laplacian is represented by a finite-difference stencil.
In order to avoid diagonalization of the full Hamiltonian, we use the iterative eigensolver ARPACK.
For each of the systems shown in Fig. 2 (a)-(c) we compute the screened interaction between charges located in the MoS$_2$ layer, and then compute the resonance eigenvalue of the complex scaled 2D Mott-Wannier Hamiltonian.

\begin{figure}
 \includegraphics[width=\columnwidth]{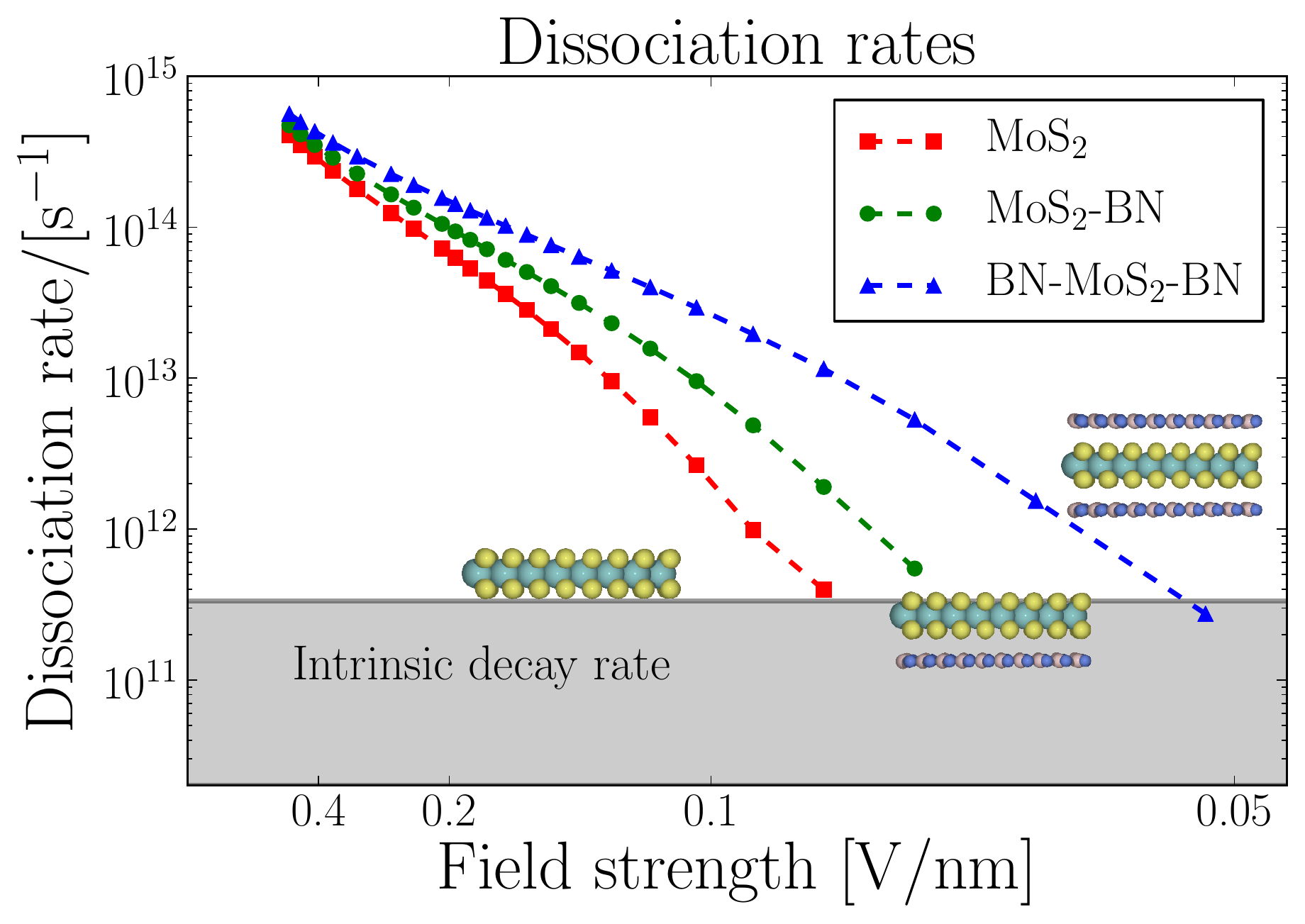}
 \caption{ \label{fig:rates-inverse} The dissociation rate of an exciton in the MoS$_2$ layer as a function of in-plane field strength for the three heterostructures.
The intrinsic decay rate spans between the defect-assisted fast decay of the excitons of 2-5 ps (upper limit) and the much slower radiative recombination of the excitons (lower limit).
 }
\end{figure}
Figure \ref{fig:rates-inverse} shows the MoS$_2$ exciton dissociation rate as a function of in-plane field strength for three different heterostructures.
As expected, larger fields lead to shorter lifetimes, and the rate is seen to depend roughly exponentially on $1/E$ for the considered field strenghts.
Furthermore, the dissociation rate can be tuned to a high degree by changing the environment of the \moly.
When \moly is placed on a single layer of boron nitride, the extra screening greatly increases the dissociation rate, and when the \moly is sandwiched between two layers of BN, the rate is even larger.
This is as expected, since larger screening results in more weakly bound excitons, which should dissociate more readily.
Adding more hBN layers on either side is expected to enhance the screening and hence the dissociation rates even further.
However, as the linear screening model breaks down in this regime\cite{latini2015a}, this has not been pursued here.

In a real device, the field-induced dissociation of excitons described here is in competition with other decay mechanisms, such as direct radiative recombination\cite{palummo2015a}. defect-assisted recombination\cite{Shi2013} and exciton-exciton annihilation\cite{Sun2014}.
The relative importance of these effects is highly dependent on the temperature of the \moly, the concentration of defects and the exciton density.

At very low temperatures, the direct radiative decay of zero momentum excitons dominates, with a characteristic lifetime of $\sim $200 fs\cite{palummo2015a,wang2016a,poellmann2015a}.
At room temperature, most of the excitons have non-vanishing momenta, and the radiative recombination lifetime is $\sim 1$ ns\cite{Shi2013, palummo2015a}.
For these systems, defect-assisted recombination therefore becomes an important mechanism, with a characteristic lifetime of 2-5 ps\cite{Shi2013, Korn2011, lagarde2014a}.
Exciton-exciton annihilations become important only when the density of excitons in a sample is large; equivalently when the average distance between excitons is small. 
At a density of $1\times 10^{12}$ cm$^{-2}$, the effective lifetime from annihilation is 11 ps\cite{Sun2014}.

Our calculations indicate that for field strengths larger than 0.1 V/nm, the dissociation lifetime is shorter than 1 picosecond in all the systems considered. This is shorter than the smallest characteristic lifetimes of the alternative decay channels at room temperature (indicated by the gray shaded region in Fig. \ref{fig:rates-inverse}) and thus field-induced dissociation should dominate. We note that a potential gradient of 0.1 V/nm is not unlikely to exist in the metal-MoS$_2$ contact region where charge transfer and interface dipole formation driven by Fermi level mismatch can lead to significant variations in the potential and band energies even in the absence of an applied bias voltage. Recently, a chemical treatment has successfully been used to eliminate the influence of defects on exciton decay\cite{amani2015a}, leaving radiative decay as the main decay channel and leading to exciton lifetimes of 10 ns. In this case, field-induced dissociation should be the main photocurrent generation mechanism.

In summary we have used complex scaling to compute the lifetime of excitons in two-dimensional MoS$_2$ and MoS$_2$/hBN structures under an applied static electric field.
The exciton was simulated using a 2D Mott-Wannier model which has previously been found to yield a reliable description of the lowest lying excitonic states in transition metal dichalcogenides.
We found that for field strengths above 0.1 V/nm, the exciton dissociation rate is larger than the intrinsic exciton decay rate in MoS$_2$ at room temperature. Moreover, encapsulation in a few layers of hBN increases the dissociation rate by an order of magnitude for fixed field strength due to the increased screening provided by the electrons in the hBN.

The Center for Nanostructured Graphene (CNG) is sponsored by the Danish National Research Foundation, Project DNRF58.
%
\end{document}